\documentclass[preprintnumbers,amsmath,amssymb,floatfix,11pt,prd,onecolumn,
superscriptaddress,nofootinbib]{revtex4}
\usepackage{latexsym}
\usepackage{epsfig}
\usepackage{epstopdf}
\usepackage{amssymb}

\begin{document}

\title{\bf Particle Dynamics Around Weakly Magnetized Riessner-Nordstr\"{o}m Black Hole
}
\author{Bushra Majeed}
\email{bushra.majeed@sns.nust.edu.pk}\affiliation{School of Natural
Sciences (SNS), Department of Mathematics, National University of Sciences and Technology
(NUST), H-12, Islamabad, Pakistan}

\author{Mubasher Jamil}
\email{mjamil@sns.nust.edu.pk}\affiliation{School of Natural Sciences (SNS), Department of Mathematics,
National University of Sciences and Technology (NUST), H-12,
Islamabad, Pakistan}

\author{ Saqib Hussain}
\affiliation{School of Natural
Sciences (SNS), Department of Physics,  National University of Sciences and Technology
(NUST), H-12, Islamabad, Pakistan}
\begin{abstract}
{\bf Abstract:} Considering the geometry of Reissner-Nordstr\"{o}m (RN) black hole
immersed in magnetic field we have studied the dynamics of neutral and charged particles. A collision of particles in the inner stable circular orbit is considered and the conditions for the escape of colliding particles from the vicinity of black hole are given. The trajectories of the escaping  particle are discussed. Also the velocity required for this escape is calculated. It is observed that there are more than one stable regions if magnetic field is present in the accretion disk of black hole so the stability of ISCO increases in the presence of magnetic field. Effect of magnetic field on the angular motion of neutral and charged particles is observed graphically.
\end{abstract}
\maketitle
\newpage
\section{Introduction}
The geometrical structure of  the spacetime around black hole (BH) could be understood better by studying the dynamics of particles in the vicinity of black holes
\cite{frolov1998, sh5979}. 
Circular
geodesics give information about geometry of spacetime \cite{po9064, am5968}. The motion of test
particles helps to study the gravitational fields of objects
experimentally and to compare the observations with the predictions
about observable effects (light like deflection, gravitational
time delay and perihelion shift). Presence of plasma is responsible for magnetic field \cite{bo2013}, in
the the surrounding of the black hole \cite{mc2307, do2008}. Near the event horizon effect of magnetic field is strong but not enough to disturb the geometry of the black hole. Motion of a charged around black hole in the presence of magnetic field  gets influenced \cite{zn7076,bl3377}. Black holes with such scenario are known as `weakly magnetized' \cite{fr2012} i.e. the magnetic field
strength lies between $B\sim 10^4-10^8\ll 10^{19}$ Gauss. Magnetic field is responsible for transferring
energy to the particles moving in the geometry around black hole, so that their  escape to spatial infinity is possible
\cite{ko8802}. Hence the collision of charged particles near the
black hole may produce much higher energy in the presence of
magnetic field than in its absence.
 In \cite{mi4199, te0903, saqib, ja2415} the effects of magnetic fields on the charged particles around black
holes were investigated. Timelike geodesics in modified gravity black hole in the presence of axially symmetric magnetic field are studied in \cite{saqib-mog}. In \cite{gulmina} authors have studied the dynamics of a charged particle around a weakly magnetized naked singularity, in the Janis-Newman-Winicour (JNW) spacetime. Kaya \cite{ka1107} studied the motion of
charged particles around a five dimensional rotating black hole in a
uniform magnetic field and found stable circular orbits around the
black hole.
In literature many aspects of the particles motion in the vicinity
of RN-black hole have been studied. In \cite{za7110} authors have studied the
high energy collisions phenomenon between the particles, currently
termed as BSW mechanism. In \cite{pu5211} the spatial regions for circular motion of neutral and
charged test particles around RN-BH and naked singularities have been studied.
Critical escape velocity for a charged particle moving around a weakly magnetized Schwarzschild black hole has been studied in \cite{za4313}. 
We consider Reissner-Nordstr\"{o}m (RN) black hole is surrounded by an
axially-symmetric magnetic field which is homogeneous at infinity.
Particles in the accretion disc moves in circular orbits in the
equatorial plane. Following the work done by Zahrani $\text{et. al.}$ \cite{za4313}, collision of a neutral and a charged particle with
another neutral particle is studied in the vicinity of magnetized
RN-BH. We focus under what circumstances the particle can escape to
infinity after collision? To evade the complication in modeling
the particle's motion around a black hole under the influence of both
gravitational and magnetic forces, we first consider the motion of a
neutral particle in the absence of magnetic field. The
study of particle dynamics around RN spacetime is also relevant as
this metric represents the extreme RN-BH and a naked
singularity as special cases.

Outline of the paper is as follows: In section II metric of RN-BH is discussed  and escape
velocity of a neutral particle is calculated. In section III the equations of
motion of a charged particle moving around weakly magnetized RN-BH
are derived. Trajectories of the particles moving around the extremal
RN-BH are plotted in section IV. In section V the dimensionless form of the equations are
given. In section VI trajectories for escape energy and escape
velocity of the particle are presented. Motion of
the particle is initially considered in the equatorial plane for the sake of simplicity. The metric signature is $(-,+,+,+)$ and
$c=1,G=1$.
\section{Escape Velocity For a Neutral Particle}
We first work for the escape velocity of a neutral particle in the absence of  magnetic field. The RN-BH
metric is given by
\begin{eqnarray}\label{1}
ds^{2}&=&-f(r)dt^2+ \frac{1}{f(r)}dr^2 +r^2(d\theta^2 +\sin^2\theta
d\phi^2),
\end{eqnarray}
where
\begin{equation}
\label{metric}f(r)=1-\frac{2M}{r}+\frac{Q^2}{r^2}.
\end{equation}
Here $M$ is the mass  and $Q$ is electric charge of the black hole. Horizon of the RN-BH
are located at: \begin{equation} r_{\pm}:= M\pm\sqrt{M^2-Q^2}.\end{equation} If
$M>Q$, there are two real positive roots; the
larger root corresponds to the event horizon and the smaller root
refers to the Cauchy horizon which is associated with the timelike
singularity at $r=0$. Black hole is known to be extremal black hole if $M=Q$ and it has only one event horizon at $r=M=Q$. If $M<Q$, there
is no real root of the equation $f(r)=0$ and there is no event
horizon. This case is known as naked singularity of RN spacetime. Symmetries of the black hole metric are along the time translation and rotation around symmetry axis. The corresponding constants of motion can be calculated using the Killing vectors
\begin{equation}
\xi_{(t)}^{\mu}\partial_{\mu}=\partial_{t} , \qquad
\xi_{(\phi)}^{\mu}\partial_{\mu}=\partial_{\phi},
\end{equation}
here
$\xi_{(t)}^{\mu}=(1,~0,~0,~0)$ and $\xi_{(\phi)}^{\mu}=(0,~0,~0,~1)
$.
The corresponding conserved quantities are the total energy
$\mathcal{E}$ of the moving
particle and its azimuthal angular momentum $L_{z}$
\begin{equation}\label{2}
\dot{t}=\frac{\mathcal{E}}{f},~~~~ \dot{\phi}=\frac{L_{z}}{r^2}.
\end{equation}
Over dot is the differentiation with respect to proper time
$\tau$ and $f:=f(r)$. From the
normalization condition $u^{\mu}u_{\mu}=-1$, we have
\begin{eqnarray}\label{4}
\dot{r}^{2}=\mathcal{E}^2- U_\text{eff},~~~~~
U_\text{eff}= (1 - \frac{2 M}{r} + \frac{Q^2}{r^2})(1 +
\frac{L_z^2}{r^2}),
\end{eqnarray}
we considered $\theta=\pi/2$ , i.e. the
planar motion of the particle.
Solving $\frac{dU_\text{eff}}{dr}=0$, we get the value of $r$ corresponding to the extreme values of the effective potential (the convolution point) \cite{chandrasekher1983}:
\begin{equation}\label{e1}
r_{\pm}=\frac{L_{z}^{2}\pm\sqrt{L_{z}^{4}-L_{z}^{2}8M^{2}}}{2M}
\end{equation}
The ISCO is at $r=4M$ for extreme RN-BH $(Q=M)$. For $Q=0$ it reduces to $r=6M$ (Schwarzschild black hole) \cite{za4313}. The corresponding energy and the azimuthal
angular momentum of the particle (in ISCO) are
respectively
\begin{equation}\label{cir2}
\mathcal{E}_{o}=\frac{(Q^2 + r_{o} (-2 M+ r_{o}))^2}{r_{o}^2 (2 Q^2 + r_{o} (-3 M
+r_{o}))},~~~~~ L_{zo}= \frac{\sqrt {Q^2 r_{o}^{2} - M r_{o}^3}}{\sqrt{-2 Q^2 + 3 M r_{o} -
r_{o}^{2}}}.
\end{equation}
For extremal black hole case, i.e. at $Q=M$, Eq. (\ref{cir2})  becomes
\begin{equation}\label{ee}
\mathcal{E}^{(e)}_{o}=\frac{(M^2 + r_{o} (-2 M + r_{o}))^2}{r_{o}^2 (2 M^2 + r (-3 M
+r_{o}))},~~~~~ L^{(e)}_{zo}= \frac{\sqrt {M (M- r_{o}) r_{o}^2}}{\sqrt{-2 M^2 + 3 M r_{o} -
r_{o}^2}}.
\end{equation}
Now consider the collision of a particle, moving in the ISCO,
with another particle which is coming from infinity (initially at
rest). This collision may result in three possibilities (depending on
the progression of the collision): (i) a bounded motion (ii) particle captured
by black hole (iii) particle escape to infinity. Orbit of the particle alters slightly if the energy and
angular momentum of the particle do not undergo a major change, otherwise the particle can move away from the original
path resulting  in captured by black hole or an escape to infinity may occur. Collision of the particles changes the equatorial
plane of the moving particle but since the metric is spherical
symmetric so  all the equatorial planes are similar.  We consider the collision occurring in such a way that $(i)$ the
azimuthal angular momentum remains invariant and $(ii)$ initial radial
velocity also remains same. Hence only the change in energy will be considered for determining the motion of particle after collision. These condition are imposed for simplification only. Particles gains an escape velocity  $(v_{\text{esc}})=v_{\bot}$ in
orthogonal direction of the equatorial plane after collision
\cite{frolov3410} and its momentum and energy (in the new  equatorial plan)
become
\begin{equation}
L^{2}=r_{o}^{2}v_\text{esc}^{2}+L_{zo}^{2},
\end{equation} here $v_\text{esc}\equiv -r\dot{\theta}_{o}$ and
$\dot{\theta}_{o}$ denotes the particles's initial polar angular velocity. Energy of the particle is
\begin{equation}\label{33}
\mathcal{E}_\text{new}=\sqrt
{(1+\frac{Q^2}{r_o^2}-\frac{2M}{r_o})v_\text{esc}^2 + \mathcal{E}^2_o},~~~~~
\mathcal{E}^{(e)}_\text{new}=\sqrt
{\frac{(M - r_o)^2 v_\text{esc}^2}{r_o^2} + \mathcal{E}^2_o},
\end{equation} with  $\mathcal{E}_{o}$, given in Eq. (\ref{ee}).
After collision, particle gains greater angular momentum and energy as compared to before collision.
From Eq. (\ref{33})  it is clear that in the asymptotic limit
($r\rightarrow\infty$),
$\mathcal{E}_\text{new}\rightarrow\mathcal{E}_{o}= 1$. So for unbounded motion (escape) particle requires
$\mathcal{E}\equiv \mathcal{E}_\text{new} \geq1$ . Hence for escape
to infinity the necessary condition is 
\begin{equation}
v_\text{esc}\geq \frac{r_o\sqrt{1-\mathcal{E}_o}}{\sqrt{-Q^2 + 2 M r_o
-r^2}},~~~~~v^{(e)}_\text{esc}\geq \frac{r_o\sqrt{1 - \mathcal{E}_o}}{\sqrt{-M^2 +
2 M r_o- r_o^2}},
\end{equation} we have solved equation $(\ref{33})$ taking $\mathcal{E}_\text{new}\ge1$ and the quantities with subscript $e$ denotes the extremal black hole case.
\section{Charged Particle Around RN-BH surrounded by Magnetic Field}
The presence of magnetic field
interrupts the motion of a charged particle around black hole. To know the aftermath of this perturbation let us start with  the Lagrangian of the moving particle as
\begin{equation}\label{10}
\mathcal{L}=\frac{1}{2}g_{\mu\nu}\dot{x}^{\mu}\dot{x}^{\nu}+\frac{q}{m}A_{\mu}\dot{x}^{\mu},
\end{equation} here $m$ is mass of the particle and $q$ is the charge of particle.
The Killing vector equation
$\Box\xi^{\mu}=0$, resembles to
the Maxwell equation for $A^{\mu}$  in the Lorentz gauge
$A^{\mu}_{\ \ ;\mu}=0$ \cite{wald8074} here $\xi^{\mu}$ denotes the Killing vector and $A^{\mu}$ is the 4-potential defined as \cite{wald8074}
\begin{equation}
A^{\mu}=\frac{\mathcal{B}}{2}\xi^{\mu}_{(\phi)}-\frac{q}{2 m}\xi^{\mu}_{(t)},
\end{equation}
with $\mathcal{B}$ as the magnetic field strength given as
\begin{equation}\mathcal{B}^{\mu}=-\frac{1}{2}e^{\mu\nu\lambda\sigma}F_{\lambda\sigma}~u_{\nu},\end{equation}
and using Levi Civita symbol, $\epsilon^{\mu\nu\lambda\sigma}$, one can write
\begin{equation} e^{\mu\nu\lambda\sigma}=\frac{\epsilon^{\mu\nu\lambda\sigma}}{\sqrt{-g}},\\
\epsilon_{0123}=1,\ \ g=det(g_{\mu\nu}).\end{equation}
The Maxwell tensor, $ F_{\mu\nu}$, is defined as
\begin{equation}
F_{\mu\nu}=A_{\nu,\mu}-A_{\mu,\nu}=A_{\nu;\mu}-A_{\mu;\nu}. \end{equation} For a local observer in RN geometry,
$u^{\mu}_{0}=\frac{1}{\sqrt{f}}\xi^{\mu}_{(t)}$. The only two
components of $F_{\mu\nu}$ will survive which are $F_{10}=-F_{01}$
and $F_{13}=-F_{31}$. 

Using the Euler-Lagrange equations for the Lagrangian defined in Eq. (\ref{10}) one can get easily
\begin{equation}\label{11}
\dot{t}=\frac{\mathcal{E}}{f(r)},~~~~~
\dot{\phi}=\frac{L_{z}}{r^{2}\sin^{2}\theta}-B,
\end{equation}
here
$B\equiv\frac{q\mathcal{B}}{2m}$.
The normalization condition $(u^{\mu} u_{\mu}=-1)$ gives
\begin{equation}\label{15ab}
\mathcal{E}^2=\dot{r}^2 +r^2 f \dot{\theta}^2 - U_\text{eff},~~~~~
U_\text{eff}=f\Big[1+r^2 \sin^2 \theta (\frac{L_{z}}{r^2 \sin^2\theta}-B)^2\Big].
\end{equation}
Equation of motion of a charged particle moving in an external electromagnetic field
$F_{\mu\nu}$ satisfies:
\begin{equation}\label{geodesic}
 \ddot {x}^{\mu}+\Gamma^{\mu}_{\nu \sigma} \dot{x}^{\nu} \dot{x}^{\sigma}=\frac{q}{m}F^{\mu}_{\alpha} \dot{x}^{\alpha}.\end{equation}
Using Eqs. (\ref{geodesic}) for the metric defined in (\ref{1}), we get the
dynamical equations for $\theta$ and $r$ given as:
\begin{eqnarray}\label{13}
  \ddot{\theta}&=&\frac{-2}{r}\dot{r}\dot{\theta} +\frac{L^2_{z}\cos\theta}{r^4 \sin^3\theta}-B^2 \sin\theta \cos\theta,\end{eqnarray}
\begin{eqnarray}\label{14}
  \ddot{r}&=&\dot{\theta}^2 \Big(\frac{2 r^2 -3rr_g+4Q^2}{2r}\Big)-\frac{r_g}{2r^2}+\frac{Q^2}{r^3}+\frac{L_z^2}{2r^2
\sin^2\theta}\Big(\frac{2 r^2-3 r r_g+4Q^2}{r^3}\Big)
-B^2 \sin^2\theta \Big(\frac{2r^2-rr_g}{2r}\Big)\nonumber\\&&+BL_z\Big(\frac{rr_g-2Q^2}{r^3}\Big)+
\frac{q^2{\mathcal{E}}}{2m^2}\Big(\frac{rr_g-2Q^2}{r^3}\Big),
\end{eqnarray}
here $r_g=2M$.
\section{Dynamical Equations in Dimensionless Form}
For the sake of convenience we can rewrite the dynamical equations
of $r$ and $\theta$ in dimensionless form by introducing the
following dimensionless quantities \cite{frolov3410}:
\begin{equation}\label{dimenless}
\sigma=\frac{\tau}{r_{g}},\ \rho=\frac{r}{r_{g}},\ \ell=\frac{L_{z}}{r_{g}},\
b=Br_{g},\  \tilde{q}=\frac{Q}{r_{g}},\ q'=\frac{q}{r_g},\ m'=\frac{m}{r_g},
\end{equation}
here $r_{g}=2M$. Using the quantities defined in Eq.
(\ref{dimenless}), Eqs. (\ref{11}) and (\ref{13})-(\ref{14}) take the forms
\begin{equation}\label{11d}\frac{d\phi}{d\sigma}=\frac{(\ell-b\rho^2 \sin^2\theta)}{\rho^2\sin^2\theta}.\end{equation}

\begin{eqnarray}\label{1a}
  \frac{d^{2}\theta}{d\sigma^{2}}&=&\frac{-2}{\rho}\frac{d\rho}{d\sigma}\frac{d\theta}{d\sigma}+\frac{\ell^2\cos\theta}{\rho^4 \sin^3\theta} -b^2\sin\theta \cos\theta,\end{eqnarray}
\begin{eqnarray}\label{1bb}
\frac{d^{2}\rho}{d\sigma^{2}}&=&\frac{1}{2m^{'2}\rho^5}\Big(\mathcal{E}\rho^2 q^{'2}(-2\tilde{q}^2+\rho)\Big)+\frac{1}{2\rho^5 }\Big(2\tilde{q}^2\rho^2-4b\ell \tilde{q}^2\rho^2-\rho^3+2b\ell \rho^3+\frac{4\ell^2\tilde{q}^2-{3\ell^2\rho}+{2\ell^2\rho}}{\sin\theta}\nonumber\\&& +b^2(1-2\rho)\rho^5\sin^2\theta+\Big(\frac{d\theta}{d\sigma}\Big)^2\rho^4\Big(4\tilde{q}^2+\rho(-3+2\rho)\Big)\Big),
\end{eqnarray}
For extremal black hole case, it becomes
\begin{eqnarray}\label{1b}
\frac{d^{2}\rho}{d\sigma^{2}}^{(e)}&=&\frac{-1}{4\rho^5}\Big[(-1+2\rho)\Big(\rho^2-2b\ell\rho^2+\frac{2\ell^2-2\ell^2\rho}{\sin\theta}+2b^2\rho^5\sin^2\theta
-2\Big(\frac{d\theta}{d\sigma}\Big)^2(-1+\rho)\rho^4\Big)\Big]+\nonumber\\&& \frac{{\mathcal{E}\rho^2 q^{'2}}}{4\rho^5m^{'2}}(-1+2\rho)
\end{eqnarray}
For the particle moving in equatorial plane, Eqs. $(\ref{1bb})$ and (\ref{1b}) become
\begin{eqnarray}\label{1bbb}
\frac{d^{2}\rho}{d\sigma^{2}}&=&\frac{1}{2m^{'2}\rho^5}\Big(\mathcal{E}\rho^2 q^{'2}(-2\tilde{q}^2+\rho)\Big)+\frac{1}{2\rho^5 }\Big(2\tilde{q}^2\rho^2-4b\ell \tilde{q}^2\rho^2-\rho^3+2b\ell \rho^3+{4\ell^2\tilde{q}^2-{3\ell^2\rho}+{2\ell^2\rho}}\nonumber\\&& +b^2(1-2\rho)\rho^5\Big),
\end{eqnarray}
and
\begin{eqnarray}\label{1c}
  \frac{d^{2}\rho}{d\sigma^{2}}^{(e)}&=&\frac{-1}{4\rho^5}\Big[(-1+2\rho)\Big(\rho^2-2b\ell\rho^2+{2\ell^2-2\ell^2\rho}+2b^2\rho^5\Big)\Big]+\frac{{\mathcal{E}\rho^2 q^{'2}}}{4\rho^5m^{'2}}(-1+2\rho)
\end{eqnarray}
Using the built in command NDSolve in Mathematica $8.0$, Eq.
$(\ref{1c})$ can be solved and the behavior of the obtained interpolating function can be better understood by plotting it against $\sigma$. Using Eq. (\ref{dimenless}) for Eqs.
$(\ref{15ab})$ we obtain
\begin{eqnarray}\label{energy}
\mathcal{E}^2&=& (\frac{d\rho}{d\sigma})^2 +\rho^2
\Big(1+\frac{\tilde{q}^{2}}{\rho^{2}}-\frac{1}{\rho}\Big)(\frac{d\theta}{d\sigma})^2+U_{\text{eff}},
\end{eqnarray}
and for extreme black hole
\begin{eqnarray}\label{energy}
\mathcal{E}^{2^{(e)}}&=& (\frac{d\rho}{d\sigma})^2 +\rho^2
(\frac{2\rho+1}{2\rho})^2(\frac{d\theta}{d\sigma})^2+U_{\text{eff}}.
\end{eqnarray}
The effective potential given in Eq. (\ref{15ab}) becomes
\begin{equation}\label{eff}
U_{\text{eff}}=\Big(1+\frac{\tilde{q}^{2}}{\rho^{2}}-\frac{1}{\rho}\Big)
\Big[1+\frac{(\ell-b \rho^2 \sin^2\theta)^2}{\rho^2
\sin^2\theta}\Big],
\end{equation}
for extreme black hole it becomes
\begin{equation}\label{eff}
U^{(e)}_{\text{eff}}=(\frac{2\rho-1}{2\rho})^2 \Big[1+\frac{(\ell-b
\rho^2 \sin^2\theta)^2}{\rho^2 \sin^2\theta}\Big].
\end{equation}
For the particle moving around RN-BH in the equatorial plane, $\theta=\pi/ 2$, at radius $\rho_{o}$, Eqs. (\ref{energy})-(\ref{eff}) become
\begin{eqnarray}\label{energy1}
\mathcal{E}^2_o&=& (\frac{d\rho_o}{d\sigma})^2+U_{\text{eff}},
\end{eqnarray}and
\begin{equation} \label{21a}
U_{\text{eff}}=\Big(1+\frac{\tilde{q}^{2}}{\rho_o^{2}}-\frac{1}{\rho_o}\Big)\Big[1+\frac{(\ell-b
\rho_o^2)^2}{\rho_o^2}\Big].
\end{equation}
for extreme black hole
\begin{equation} \label{21}
{U}^{(e)}_{\text{eff}}=(\frac{2\rho_o-1}{2\rho_o})^2\Big[1+\frac{(\ell-b
\rho_o^2)^2}{\rho_o^2}\Big].
\end{equation}
Again considering the ideal scenario of collision which does not
change the azimuthal angular momentum of the particle except its
energy i.e. $\mathcal{E}_{o}\rightarrow\mathcal{E}$, defined as
\begin{equation}\label{20}
\mathcal{E}=
  \Big[\mathcal{E}_{o}+\Big(1+\frac{\tilde{q}^{2}}{\rho^{2}_o}-\frac{1}{\rho_o}\Big) v_\text{esc}^2\Big]^{1/2},
\end{equation}
for extremal black hole
\begin{equation}\label{20a}
\mathcal{E}^{(e)}=
\Big[\mathcal{E}_{o}+(\frac{2\rho_o-1}{2\rho_o})^2 v_\text{esc}^2\Big]^{1/2},
\end{equation}
where $\mathcal{E}_{o}$ is the energy defined in Eq.
(\ref{energy1}). As already mentioned that when
$\rho\rightarrow\infty$ the energy $\mathcal{E}\rightarrow1$. So for
the unbound motion the energy of the particle should be
$\mathcal{E}\geq1$. Solving equation $(\ref{20})$ at
$\mathcal{E}=1$, for escape velocity of the particle, we get
following expression
\begin{eqnarray}\label{31a}
v_\text{esc}&=& \sqrt{\frac{1-\Bigg(1+\Big(\frac{\ell}{\rho_o^{2}}-b\Big)^{2}\Bigg)^{2}
\Big(1+\frac{\tilde{q}^{2}}{\rho_o^{2}}-\frac{1}{\rho_o}\Big)^{2}}
{1+\frac{\tilde{q}^{2}}{\rho_o^{2}}-\frac{1}{\rho_o}}},
\end{eqnarray}
and for extremal RN-BH
\begin{eqnarray}\label{31}
v^{2^{(e)}}_\text{esc}&=& \frac{-\ell^2 (1-2\rho_o)^2 +2b\ell (1-2\rho_o)^2 \rho_o^2
-\rho_o^2(1+\rho_o(-4+b^2(1-2\rho_o)^2\rho_o))}{(1-2\rho_o)^2\rho_o^2}.
\end{eqnarray}
For simplicity we are considering the particle to be initially in
ISCO, further we discuss the behavior of the particle when it
escapes to asymptotic infinity. The only parameters required for
describing the motion of the particle are the parameters $\ell$ and
$b$ defined in term of $\rho_{o}$ and the energy of the particle.
The expression for the parameters $\ell$ and $b$ in term of $\rho$
could be obtained by dealing with Eq. (\ref{21}). The first and
second derivatives of the effective potential defined in Eq.
(\ref{21}) are\begin{equation}\label{22} \frac{d}{d\rho}U_\text{eff}= \frac{(-1 +
2 \rho_o) (-2\ell^2 (-1 + \rho_o) + \rho_o^2 -
2 b \ell \rho_o^2 + 2 b^2 \rho_o^5)}{2 \rho_o^5},\end{equation}
and
\begin{eqnarray}\label{23}
\frac{d^2}{d\rho^2}U_\text{eff}=\frac{2 b \ell \rho_o^2 (-3 + 4 \rho_o) +
2 \ell^2 (5 + 6 (-2 + \rho_o) \rho_o) + \rho_o^2 (3 -
4 \rho_o + 4 b^2 \rho_o^4)}{2 \rho_o^6}.
\end{eqnarray}
The values of
$\ell$ and $b$ can be found by solving simultaneously
$\frac{d}{dr}U_\text{eff}=0$ and $\frac{d^2}{dr^2} U_\text{eff}=0$, as given
below
\begin{eqnarray}\label{l}
\ell&=&\pm\frac{\sqrt{-3\rho_o^5(-2+\rho_o)^3}\sqrt{-\frac{10\rho_o^3-13\rho_o^4+4\rho_o^5+\sqrt{-3(-2+\rho_o)^3\rho_o^5}}{(1-2\rho_o)^2\rho_o^5(-3+2\rho_o)}}}{2(-2+\rho_o)^2},
\end{eqnarray}
\begin{eqnarray}\label{b}
b&=&\frac{1}{2}\sqrt{-\frac{10\rho_o^3-13\rho_o^4+4\rho_o^5+\sqrt{-3(-2+\rho_o)^3\rho_o^5}}
{\rho_o^5(-1+2\rho_o)^2
(2\rho_o-3)}}.
\end{eqnarray}
In section VII the
parameters $\ell$ and $b$ are plotted against $\rho$.
\section{Effect of Magnetic Field on Motion of Particles}
\begin{figure}[!ht]
\begin{minipage}{.45\linewidth}
\includegraphics[width=7.0cm]{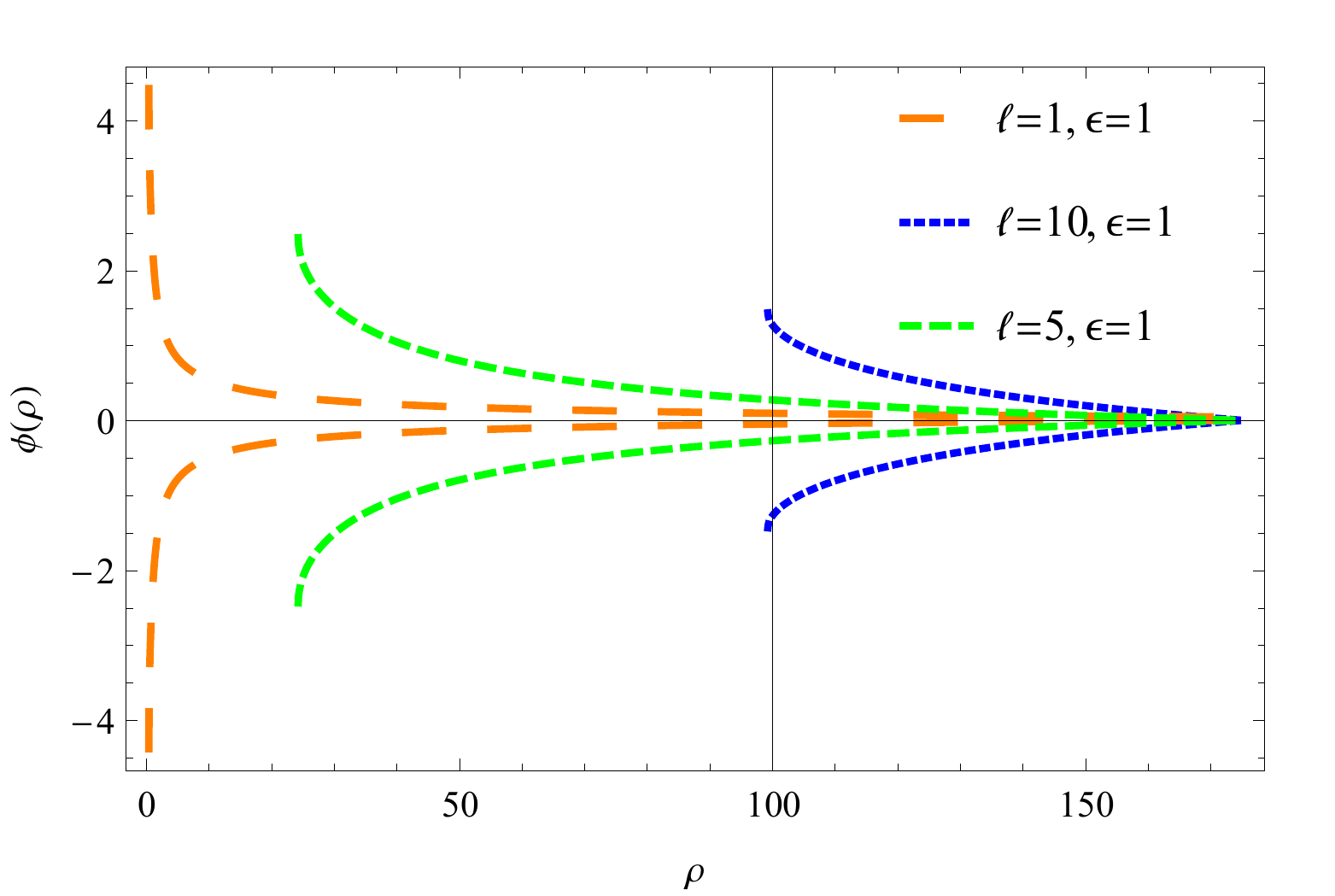}
\caption{Behavior of angular motion linked with angular momentum of the particle.}\label{v1}
\end{minipage}
\quad
\begin{minipage}{.45\linewidth}
\includegraphics[width=7.0cm]{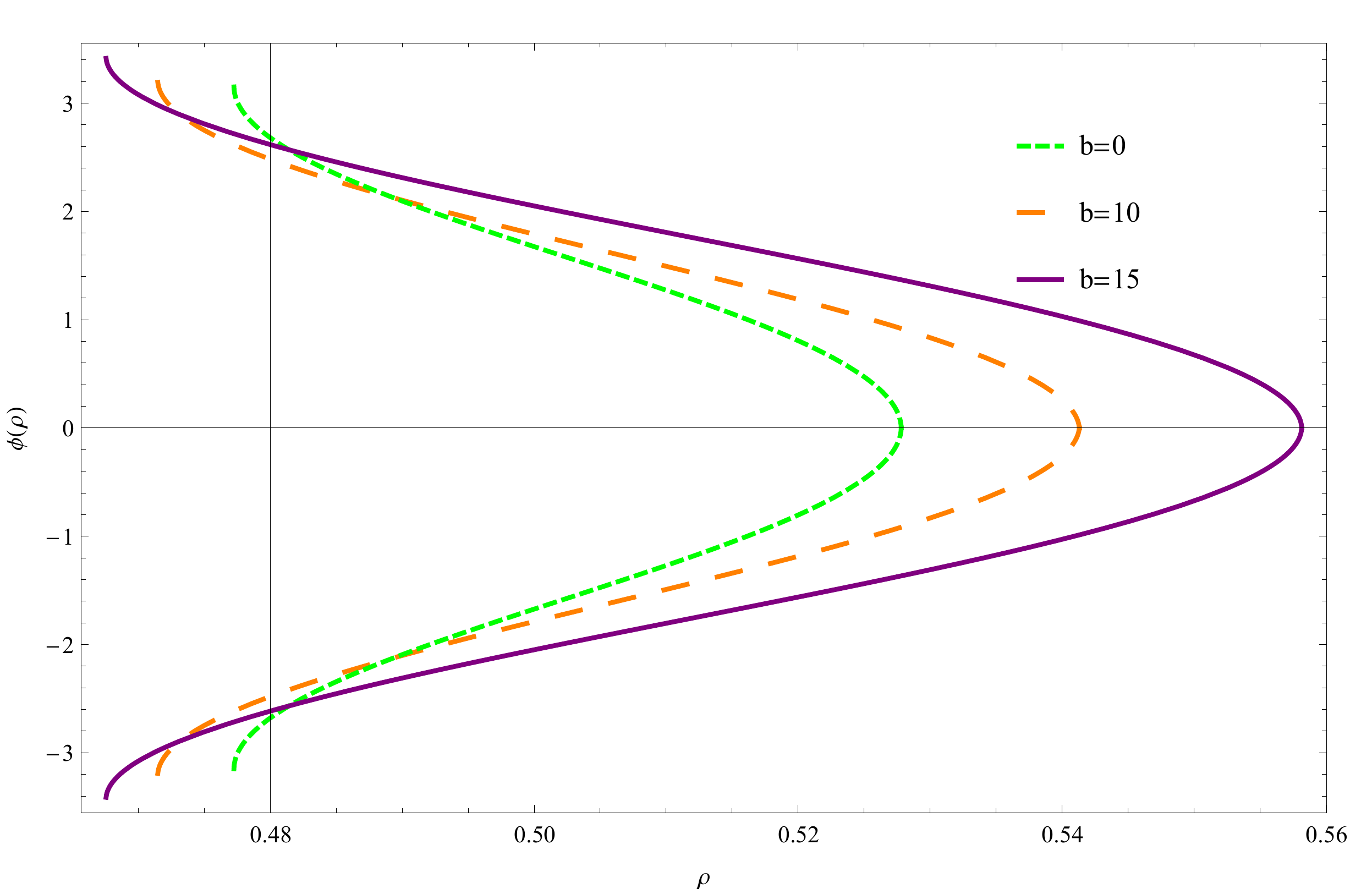}
\vspace{0.0cm}\caption{Effect of magnetic field on the charged moving particle, for $\mathcal{E}=1$, $\ell=10$}\label{v2}
\end{minipage}
\end{figure}

Consider the neutral particle moving around RN-BH. Writing  the equations associated with the constants of motion, in dimensionless form we have
\begin{equation}\label{g1}\frac{d\rho}{d\sigma}=\pm \frac{\sqrt{\rho^4 \mathcal{E}^2-(\rho^2-\rho+\tilde{q}^2)(\rho^2+\ell^2)}}{\rho^2},\end{equation}
where positive sign is for the  particle going away from
the black hole, and negative sign is for the path of an ingoing particle, also
\begin{equation}\label{g2}\frac{d\phi}{d\sigma}=\frac{\ell}{\rho^2}.\end{equation}
Using Eqs. (\ref{g1}) and (\ref{g2}) together we have
\begin{equation}\label{a}\frac{d\phi}{d\rho}=\frac{\ell}{\sqrt{\rho^4{\mathcal{E}^2}-\Big((\rho^2-\rho+\tilde{q}^2)(\rho^2+\ell^2)\Big)}}.\end{equation}
For extremal black hole it becomes
\begin{equation}\label{aa}\frac{d\phi}{d\rho}=\frac{2\ell}{\sqrt{4\rho^4{\mathcal{E}^2}-\Big((2\rho-1)^2(\rho^2+\ell^2)\Big)}}.\end{equation}
It is observed graphically  that a particle having less angular momentum approaches the black hole more closely as compared to the one having large angular momentum. This shows that when a particle does not move radially, its chances for approaching the black hole event horizon are very less, Fig. (\ref{v1}).
For a charged particle moving around RN-BH
in the presence of magnetic field, we can write
\begin{equation}\label{g11}\frac{d\rho}{d\sigma}=\pm\frac{\sqrt{\mathcal{E}^2\rho^4-(\rho^2-\rho+\tilde{q}^2)(\rho^2+(\ell-b\rho^2)^2)}}{\rho^2},\end{equation} and
\begin{equation}\label{g21}\frac{d\phi}{d\sigma}=\frac{(\ell-b\rho^2)}{\rho^2}.\end{equation}
Using Eqs. (\ref{g11}) and (\ref{g21}) together we have
\begin{equation}\label{a1}\frac{d\phi}{d\rho}=\pm\frac{(\ell-b\rho^2)}{\sqrt{\rho^4{\mathcal{E}^2}-(\rho^2-\rho+\tilde{q}^2)\Big(\rho^2+(\ell-b\rho^2)^2\Big)}}.\end{equation}
For extremal black hole
\begin{equation}\label{aa1}\frac{d\phi}{d\rho}=\pm\frac{2(\ell-b\rho^2)}{\sqrt{4\rho^4{\mathcal{E}^2}-(2\rho-1)^2\Big(\rho^2+(\ell-b\rho^2)^2\Big)}}.\end{equation}
Change in  $\phi$ during the motion of
particle, around RN-BH, starting its motion from some finite distance is shown in Fig. (\ref{v2}). It is observed that behavior of angular motion of particle is linked with the strength of magnetic field.
\section{Behavior of Effective Potential}
\begin{figure}[!ht]
\begin{minipage}{.45 \linewidth}
\vspace{0.0cm}\includegraphics[width=7.0cm]{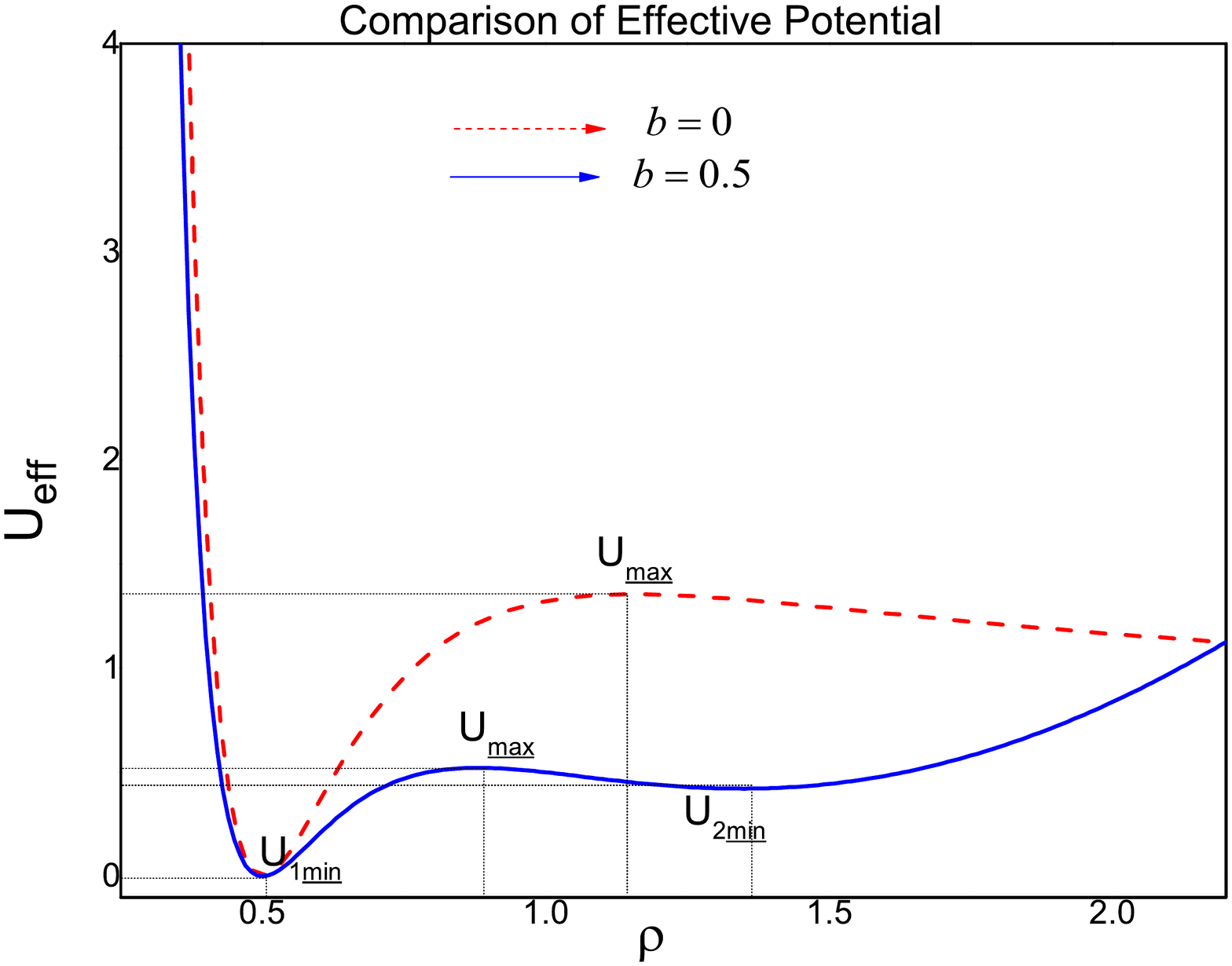}
\caption{Effective potentials (with magnetic
field, and in its absence, a comparison) versus $\rho$.}\label{f111}\end{minipage}
\quad
\begin{minipage}{.45\linewidth}
\vspace{0.0cm}\includegraphics[width=7.0cm]{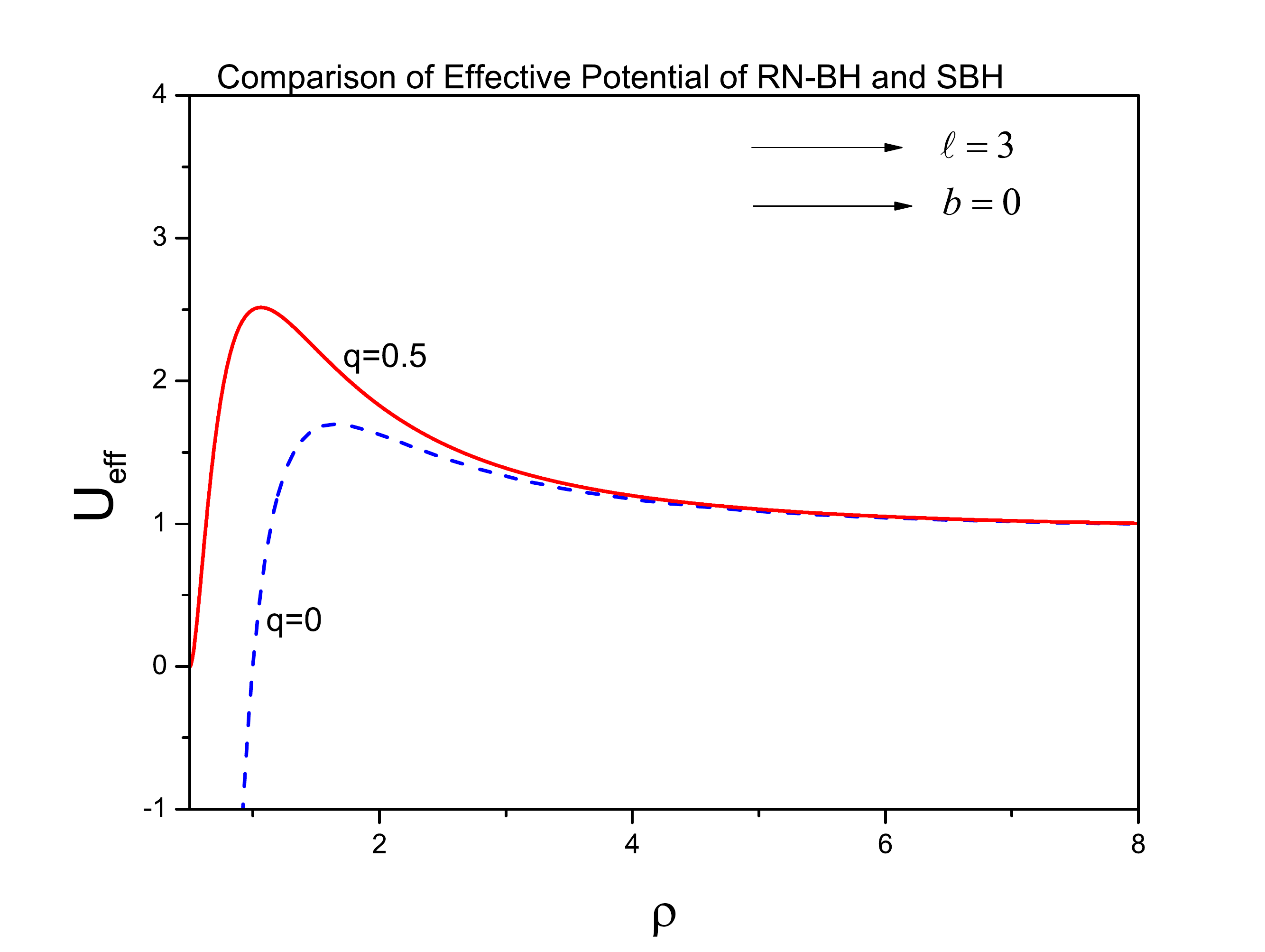}
\caption{Effective potentials Eq. (\ref{21a}) as a function of $\rho$.} \label{RSC}
\end{minipage}
\end{figure}

\begin{figure}[!ht]
\begin{minipage}{.45\linewidth}
\includegraphics[width=7.0cm]{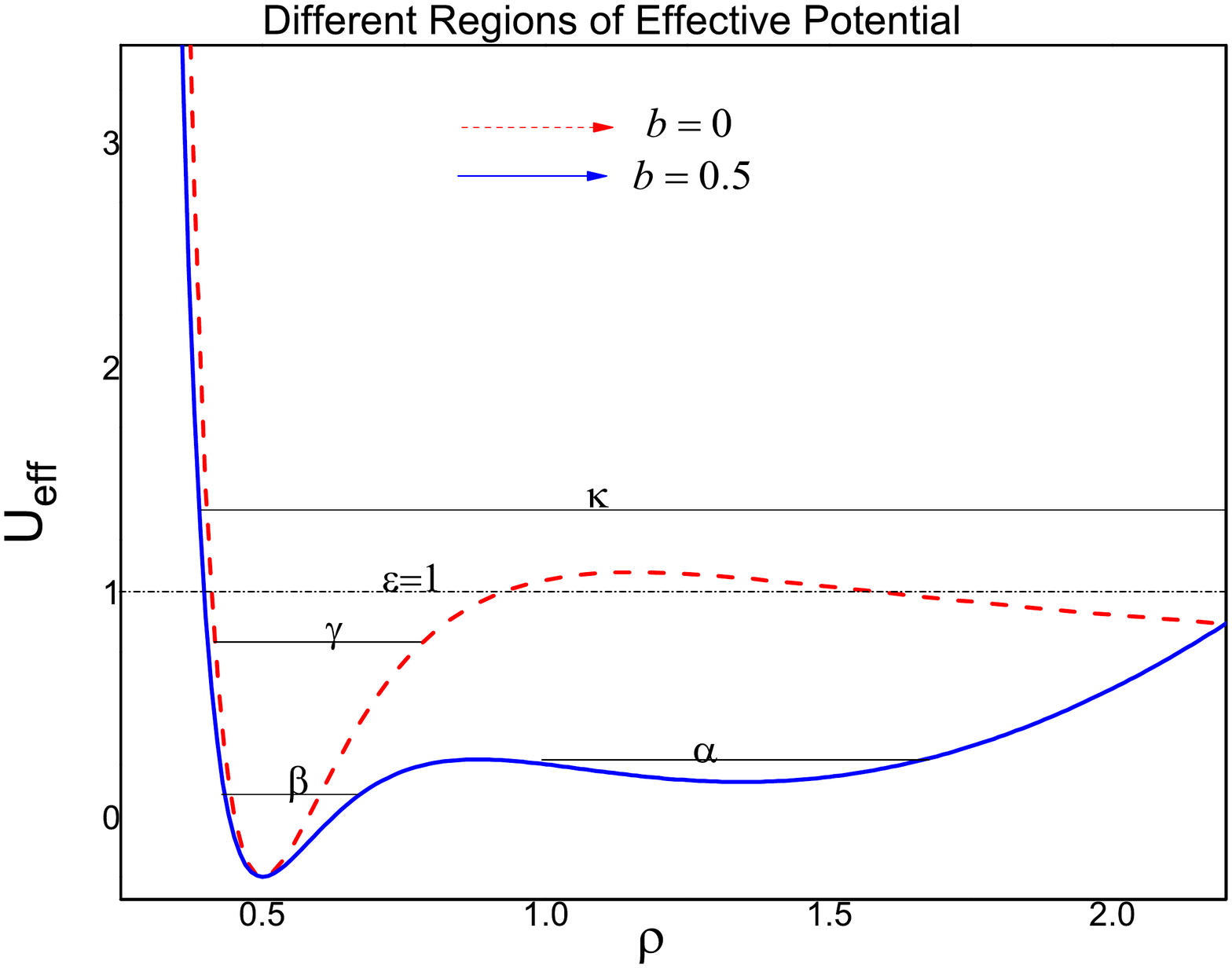}
\caption{$\alpha$
and $\beta$ are the regions which correspond to stable orbits for
$b=0.5$. For $b=0$ the region $\gamma$ corresponds to stable orbits.}\label{f2}
\end{minipage}
\quad
\begin{minipage}{.45\linewidth}
\includegraphics[width=7.0cm]{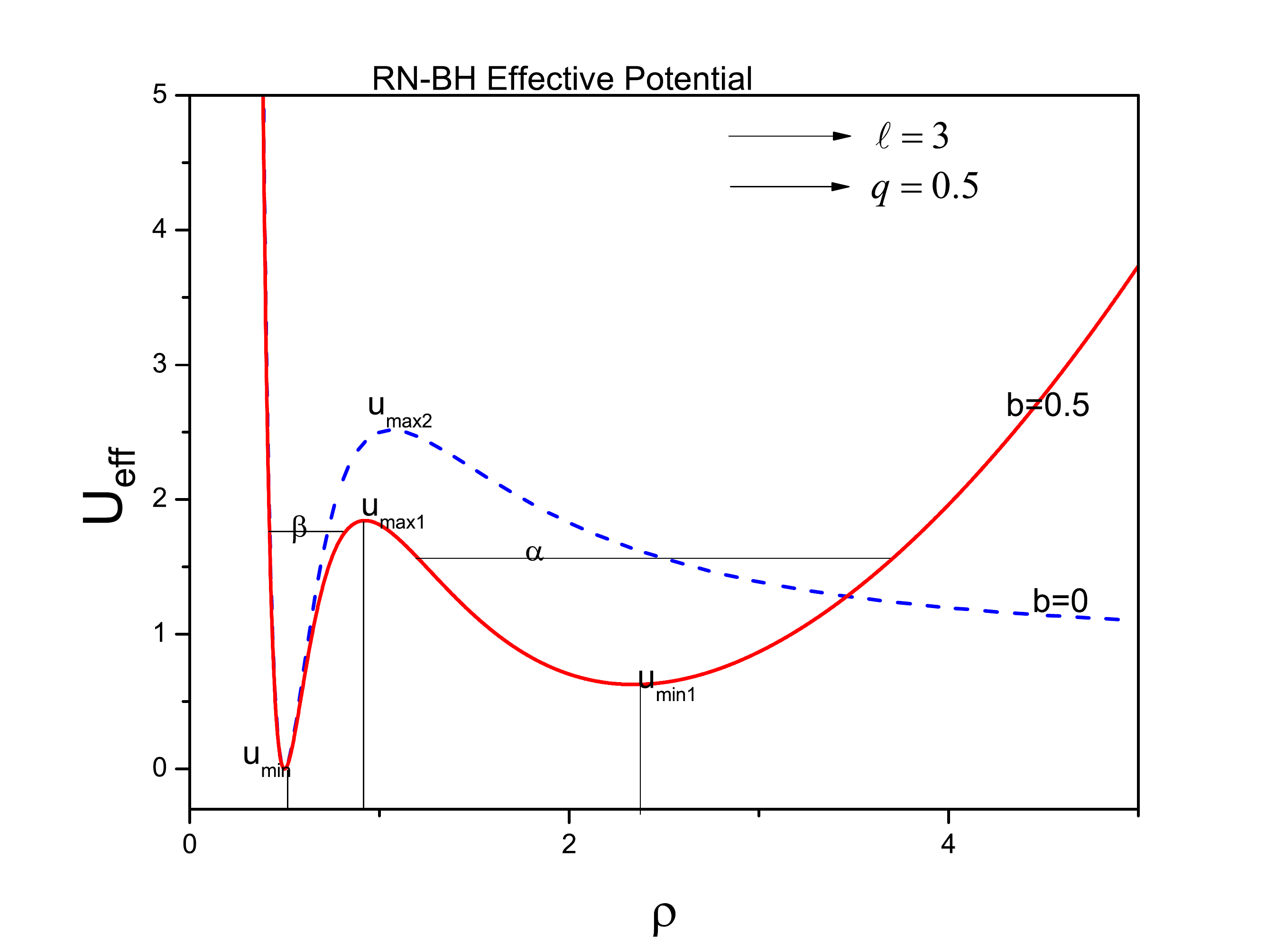}
\caption{Behavior of effective potentials Eq. (\ref{21a}) with and without magnetic
field vs $\rho$.}\label{f1}
\end{minipage}
\end{figure}

\begin{figure}[!ht]
\begin{minipage}{.45\linewidth}
\includegraphics[width=7.0cm]{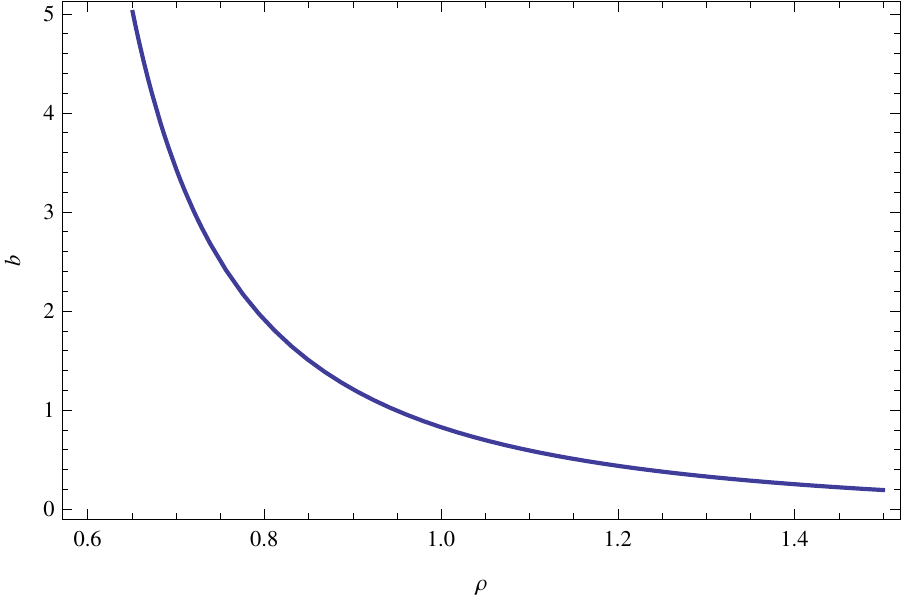}
\caption{Magnetic field $b$ as a
function of $\rho$. One can notice that magnetic field
decreases abruptly  away from the black hole.}\label{f5}
\end{minipage}
\quad
\begin{minipage}{.45\linewidth}
\vspace{-1.1cm}\includegraphics[width=7.0cm]{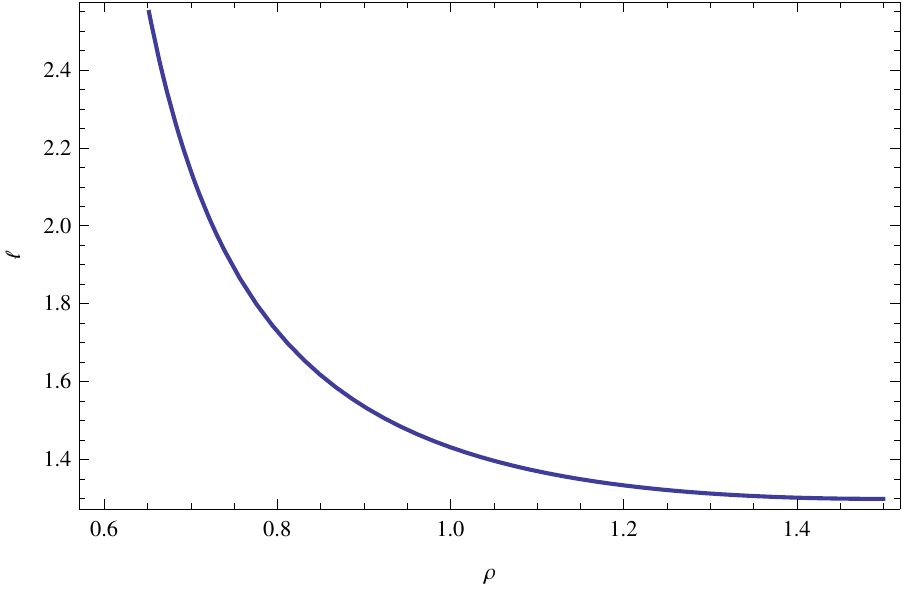}
\caption{Angular momentum as a function of $\rho$.}
\label{f6}\end{minipage}
\end{figure}

In this section the behaviors of effective potentials of particle are studied
graphically and the energy required for its escape to infinity or for bounded motion is discussed.
In Fig. (\ref{f111}) we have plotted the effective potential as a function
of $\rho$. There are two minima $u_{min}$ and $u_{min1}$,
in the presence of magnetic field while in the absence of magnetic field there is
only one minima, $u_{min}$. Hence we can say that the presence of
magnetic filed increases the possibility of the particle to move in
a stable orbit. A comparison of effective potential of
Schwarzschild black hole  with that of RN-BH is established in Fig.
(\ref{RSC}). It is clear that the maxima for the effective
potential of particle moving around RN-BH has greater value in comparison with the maxima
of effective potential for Schwarzschild black hole. Since a particle moving
around black hole could be captured if it has energy greater then maxima of
its effective potential otherwise it will move back to infinity or
may reside in some stable orbit. Therefore, we can say that the
possibility for a particle to escape from the vicinity of RN-BH or
to stay in some stable orbit is more as compared to its behavior while moving around Schwarzschild black hole. In Fig. (\ref{f2}) different regions of effective potential which
are linked to escape and bounded motion of the particle are shown.
Here $\alpha$ and $\beta$ are the regions which correspond to stable
orbits for $b=0.5$. For $b=0$ there is only one stable region
represented by $\gamma$ which is related to a stable orbits. Dotted
line represents the minimum energy required to escape from the
vicinity of black hole. If the particle has energy $\mathcal{E}\geq1$ and
move toward the black hole it will bounce back to infinity which is
represented by $\kappa$.
In Fig. (\ref{f1}) we are comparing the effective potentials of
extremal black hole in the presence of magnetic field and without magnetic
field. One can notice that for $b=0.5$, the effective potential has
two local minima which corresponds to two stable regions while for
$b=0$ it has only one minima. We use the notation $U_{max}$ and
$U_{min}$ for unstable and stable circular orbits of the particle
respectively. Here $U_{1min}$ corresponds to ISCO which coincides
with ISCO of the case when $b=0$ (dotted curve) and $U_{2min}$
correspond to stable circular orbits which occur due to presence of
magnetic field. Therefore we can say that magnetic field contributes
to increase the stability of the orbits.
In Fig. (\ref{f5}) we have plotted the magnetic field as a function
of $\rho$. One can notice that magnetic
field decreases abruptly as particle moves away from the source
(black hole). We have plotted the angular momentum $\ell$ as a function of
$\rho$ in Fig. (\ref{f6}), it is clear that
$\ell\rightarrow\infty$ for $\rho=0.5$. Angular momentum
$\ell_{\pm}$ for ISCO as a function of magnetic field $b$ are
plotted in Fig. (\ref{f6}). 
\begin{figure}
\includegraphics[width=10.0 cm]{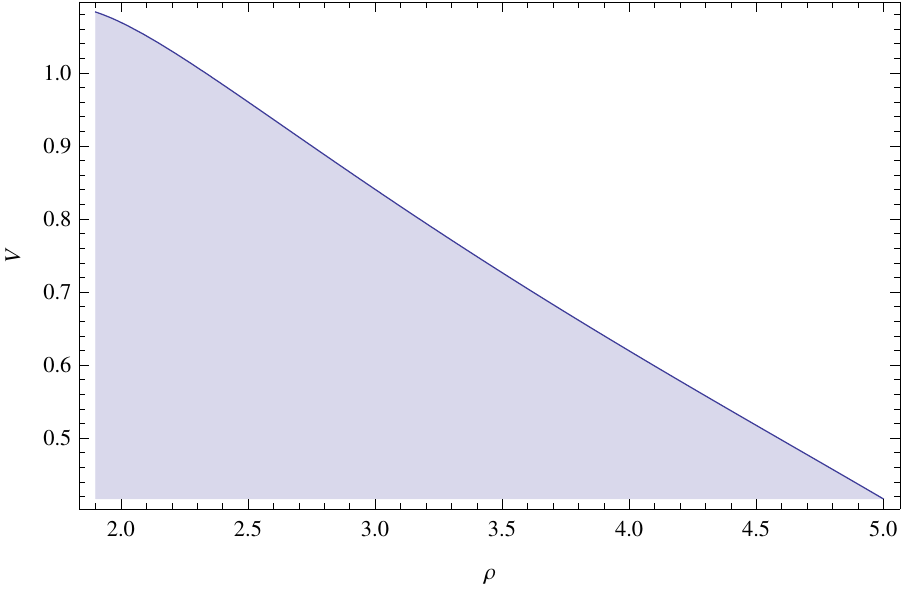}
\caption{Solid curved line
represents the minimum velocity required to escape from the
ISCO. We have plotted the escape velocity ($v_\text{esc}$) for
$b=0.9$ and $\ell=1.5$.} \label{f10}
\end{figure}
\begin{figure}[!ht]
\begin{minipage}{.45\linewidth}
\vspace{0.0cm}
\includegraphics[width=7cm]{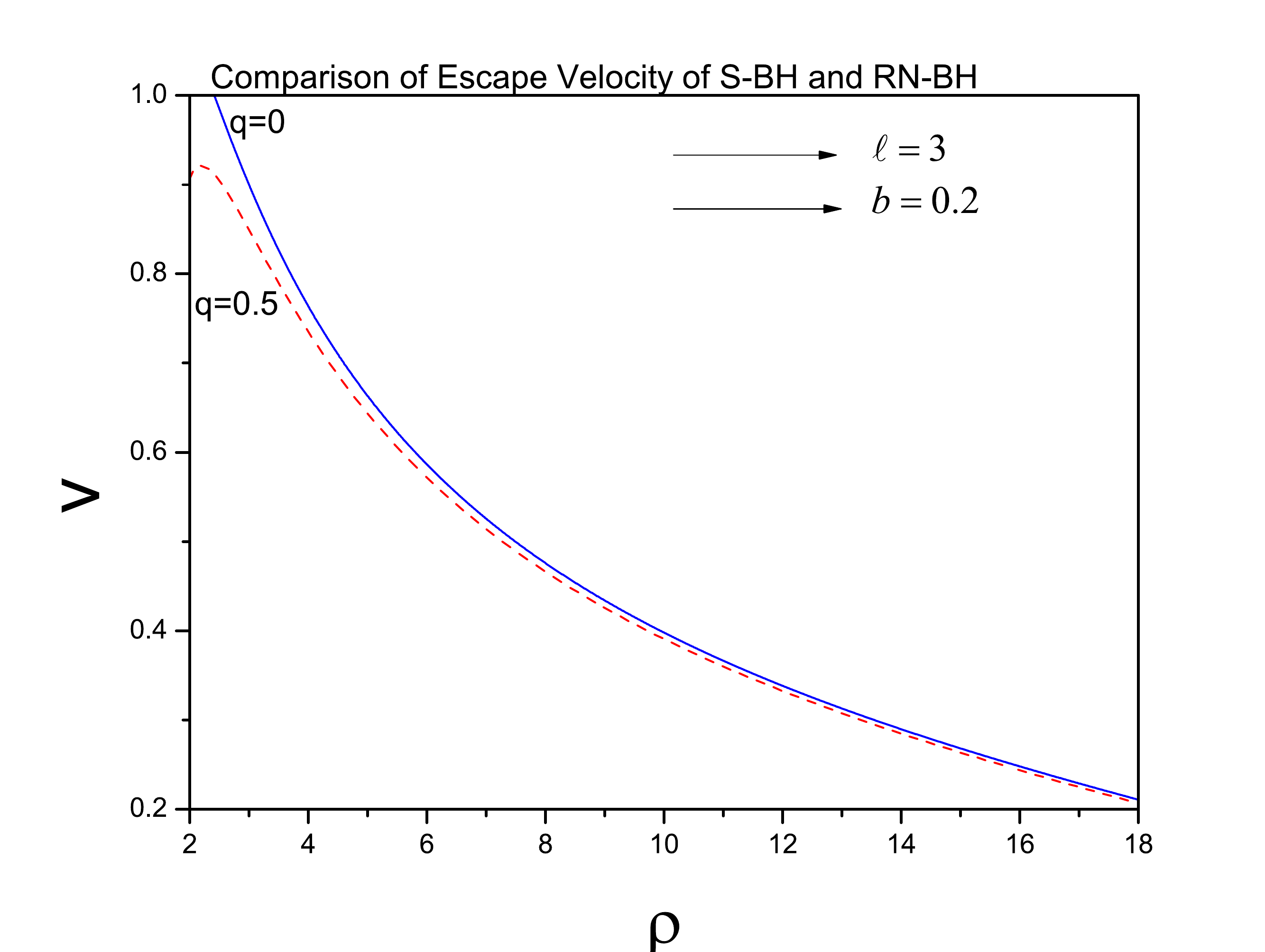}
\caption{Escape velocity as function of $\rho$
for both schwarzschild black hole and RN-BH.} \label{f7}
\end{minipage}
\quad
\begin{minipage}{.45\linewidth}
\vspace{0.5cm}
\includegraphics[width=7cm]{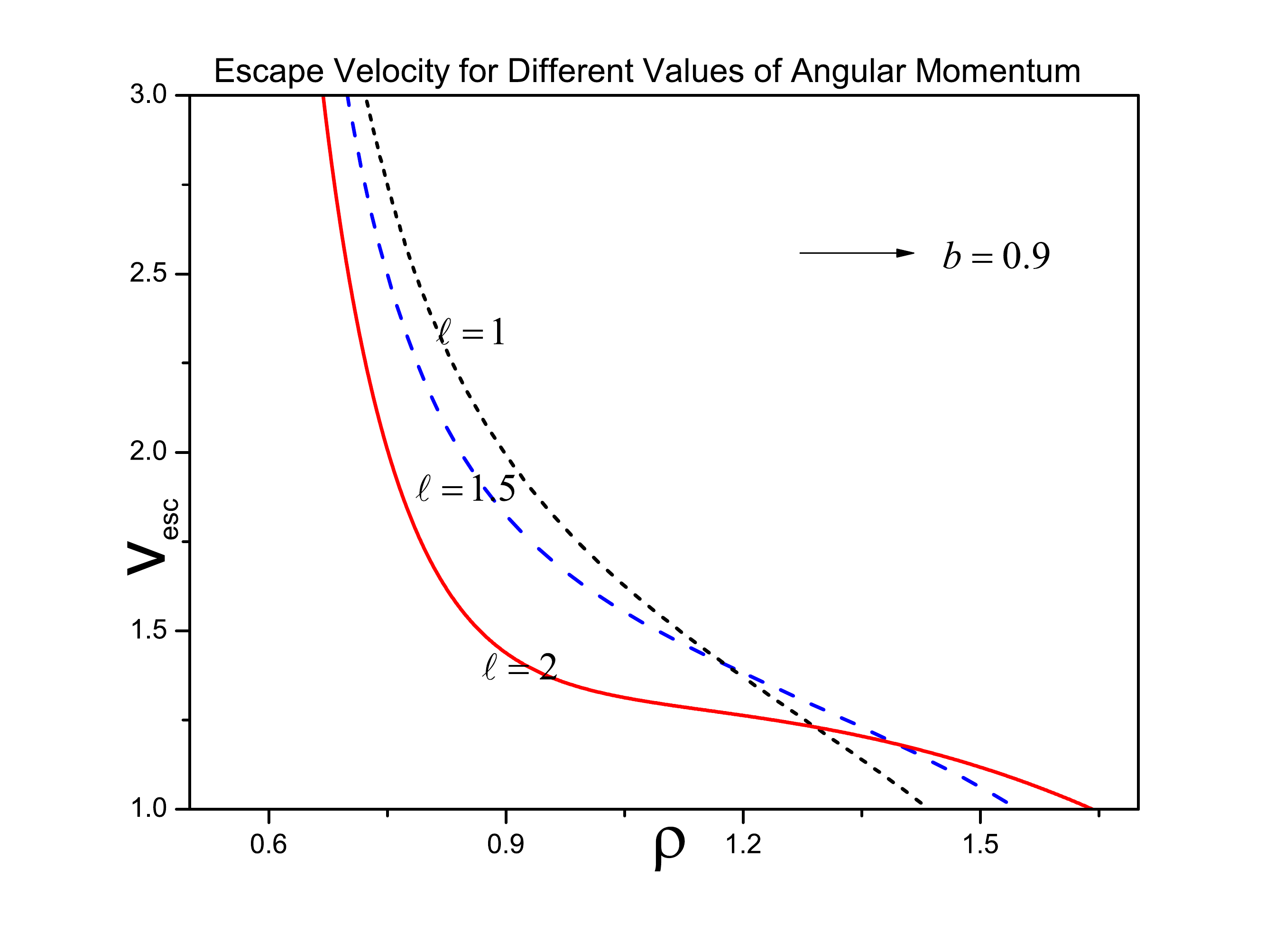}
\caption{Escape velocity ($v_\text{esc}$) against $\rho$ for different
values of angular momentum $\ell$, for extremal RN-BH.} \label{f91}
\end{minipage}
\end{figure}

\begin{figure}
\begin{minipage}{.45\linewidth}
\vspace{0.0cm}\includegraphics[width=9cm]{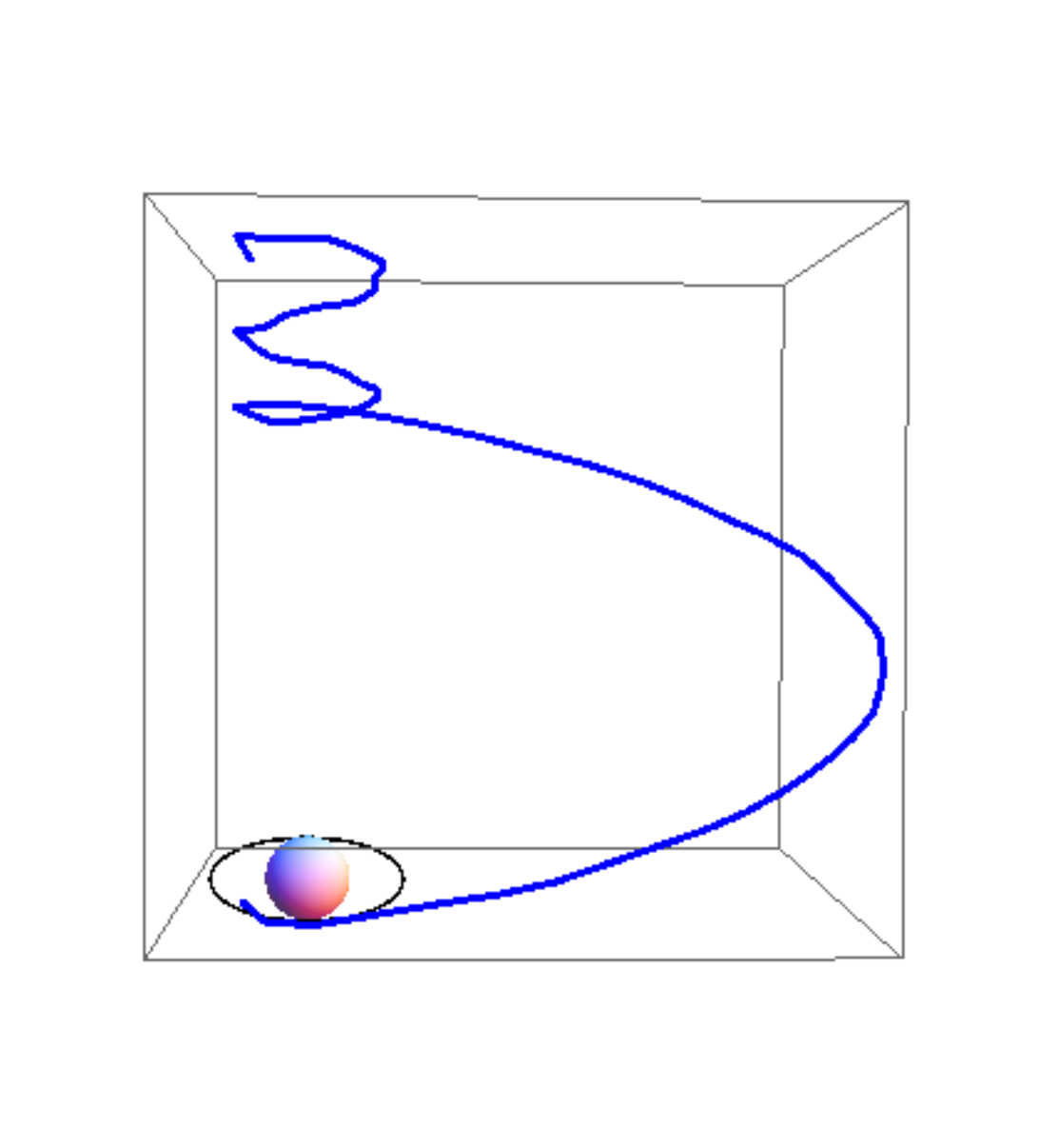}
\caption{Trajectory of the escaping particle around RN-BH, for $b=1.5$, $\ell=1$, $\mathcal{E}=2$, $\rho[0]=1$.}
\label{case6}
\end{minipage}
\quad
\begin{minipage}{.45 \linewidth}
\vspace{0.95cm}
\includegraphics[width=9cm]{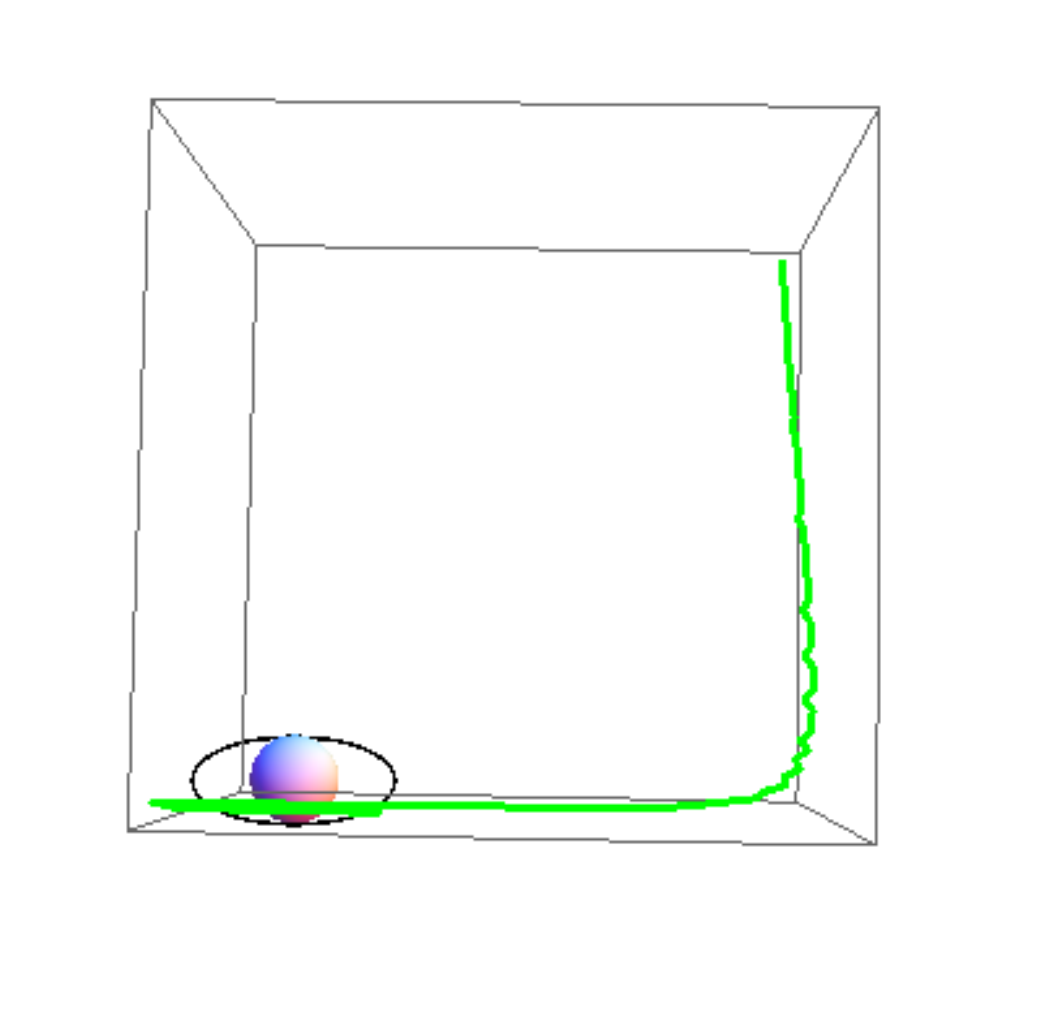}
\caption{Trajectory of the escaping particle around RN-BH, for $b=1$, $\ell=1$, $\mathcal{E}=1$, $\rho[0]=1.2$.} \label{case4}
\end{minipage}
\end{figure}
\begin{figure}\includegraphics[width=9cm]{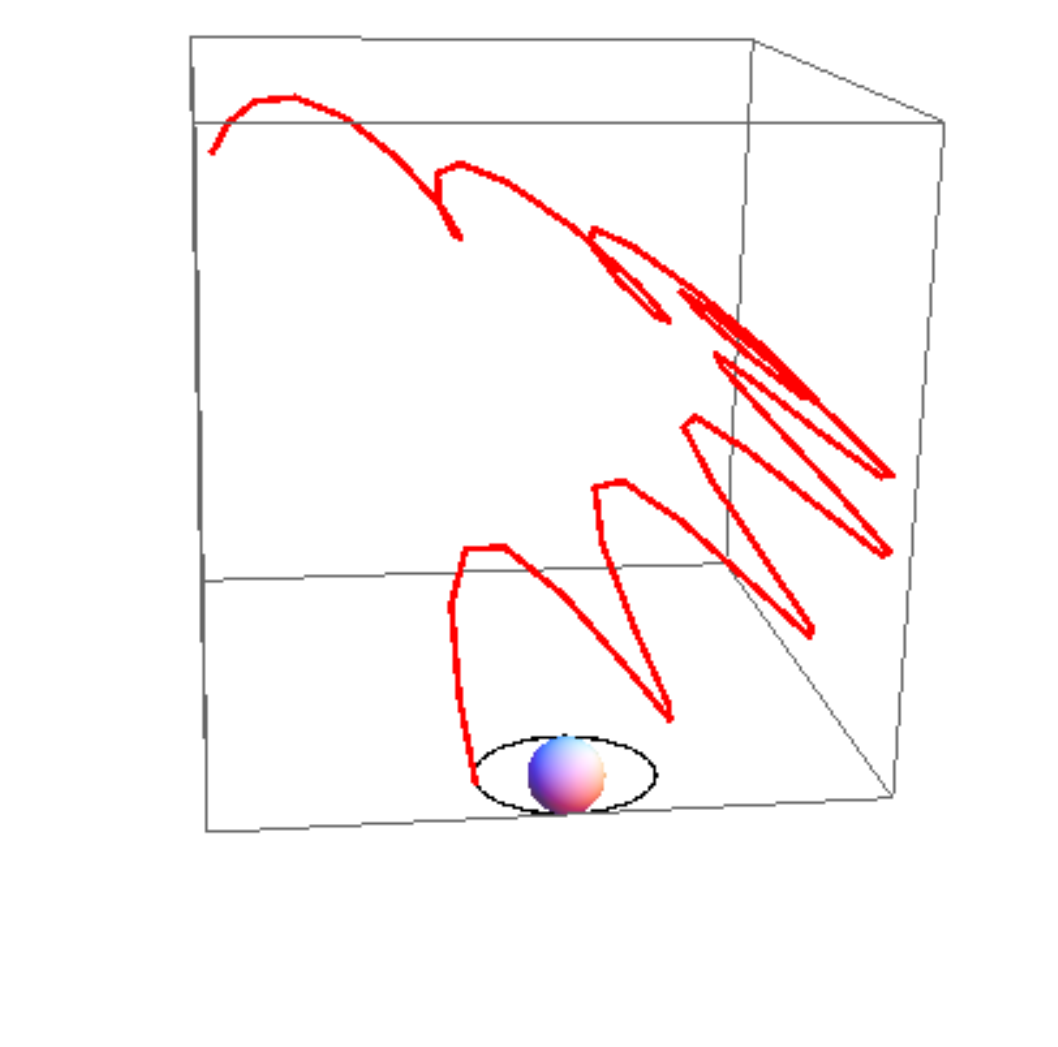}
\caption{Trajectory of the escaping particle around RN-BH, for $b=0.5$, $\ell=1$, $\mathcal{E}=1$, $\rho[0]=0.51$.}
 \label{case21}
\end{figure}
\begin{figure}[!ht]
\begin{minipage}{.45\linewidth}
\vspace{0.0cm}
\includegraphics[width=7cm]{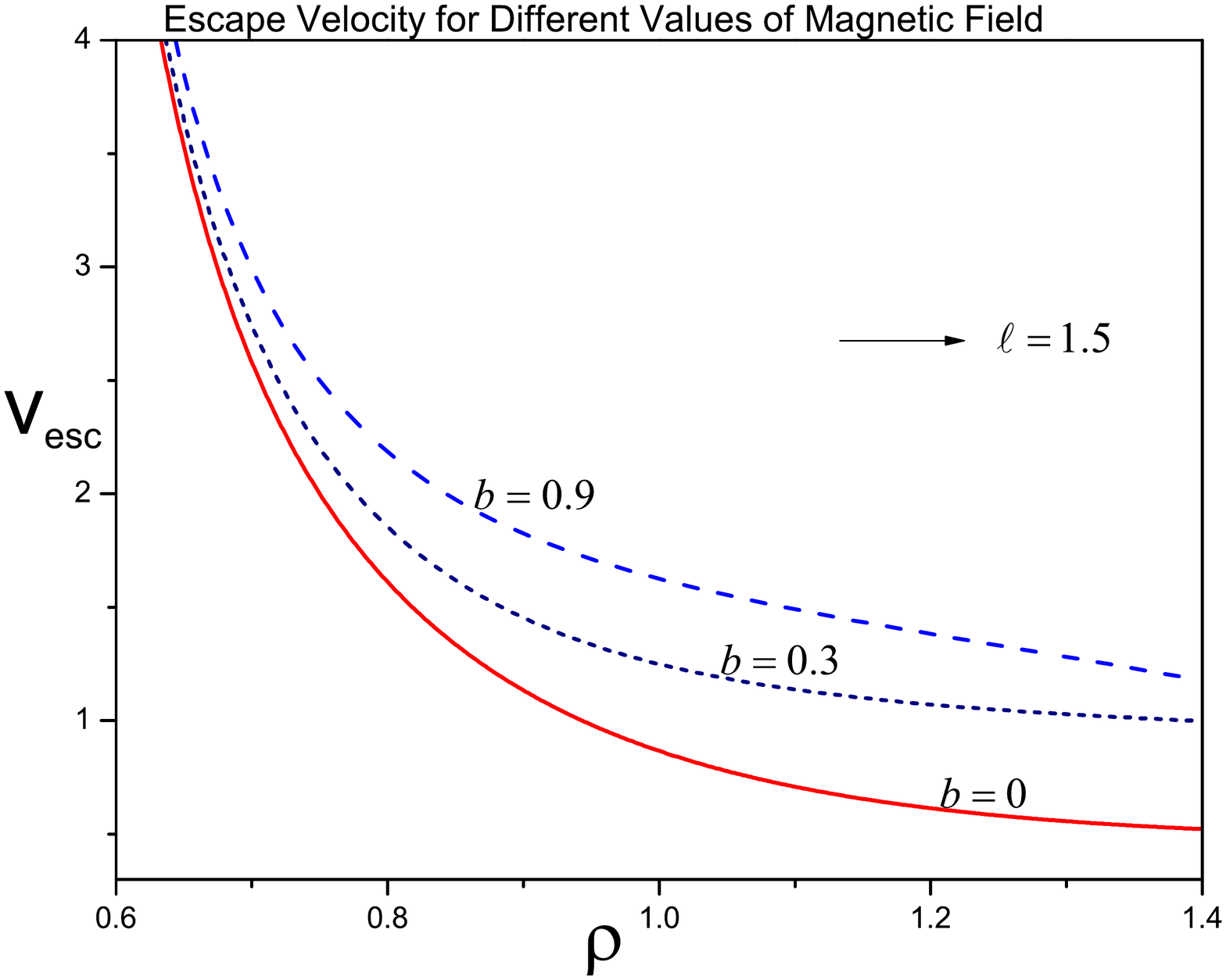}
\caption{Effect of magnetic field $b$ on escape velocity ($v_\text{esc}$).} \label{f8}
\end{minipage}
\quad
\begin{minipage}{.45 \linewidth}
\includegraphics[width=7cm]{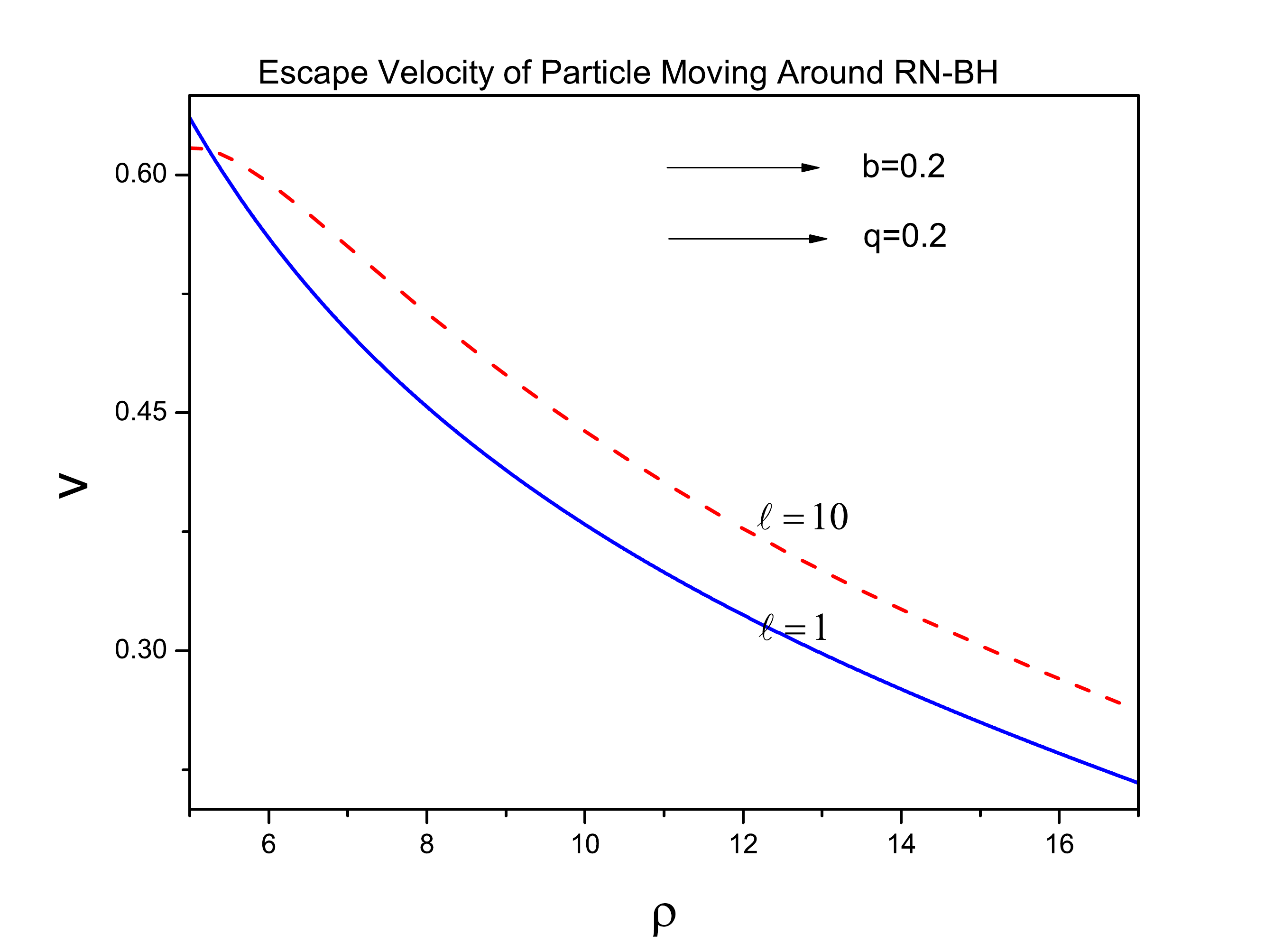}
\caption{Escape velocity ($v_\text{esc}$) against $\rho$ for different
values of angular momentum $\ell$.} \label{f9}
\end{minipage}
\end{figure}
\section{Trajectories of Escape Velocity}
Escape velocity of the charged particle is plotted in Fig. (\ref{f10}), the shaded region
correspond to escape of particle from ISCO to
$\infty$ and $-\infty$ respectively. The solid curve represents the
minimum velocity which is required to escape from the ISCO. The
unshaded region represents that if the value of  velocity lies in
this region then particle  will remain in the ISCO or some other
stable orbit. In Fig. (\ref{f7}) we are comparing the escape
velocity of a particle moving around the Schwarzschild black hole  with the
particle moving around the RN-BH. It is clear
that the difference between the velocities is large near the black hole and it become almost same as we move away from the black hole. Therefore, we can conclude that the effect of the charge
of black hole on the motion of the particle is large and it is
reducing as particle moves away from it. The behavior of escape
velocity for extreme RN-BH is shown in Fig. (\ref{f91}). The paths followed by the kicked (escaping) particle, moving  initially in the
ISCO, are shown in Figs. (\ref{case6}), (\ref{case4}), and (\ref{case21}), which are obtained by solving Eqs. (\ref{11d}), (\ref{1a}) and (\ref{1bb}) numerically,  we have taken initial  radial velocity after collision as zero. We are interested to know the effect of magnetic field on the motion of charged particle (the magnitude of deformation produced in oscillatory motion). This effect increases as strength of magnetic field increases. Escape velocity is  plotted in Fig. (\ref{f8}) for different values of magnetic field $b$. Escape of the particle from the vicinity of black hole becomes easier in the presence of stronger  magnetic field. As particle goes away from the black hole its escape velocity becomes
almost constant, just like the strength of magnetic field. Hence presence of magnetic field will
provide more energy to particle, so that it might easily escape from
the vicinity of black hole.
In Fig. (\ref{f9}) connection of  escape velocity
with angular momentum is shown. It is clear that escape velocity of a
particle with larger value of $\ell$ is greater compared to
the particle with smaller value of $\ell$.
\section{Summary and Conclusion}
Motion of particles in the RN geometry in the presence of magnetic
field is investigated in this paper. To avoid complications in the
analysis some assumption are made, as mentioned in sec II. We first
studied the neutral particle moving around RN-BH and derived the
expressions for the energy and azimuthal angular momentum of the
particle corresponding to ISCO. We obtained the expressions for
escape velocity of the particle, after its collision with some other
particle. Then analysis for a charged particle is done, and
dynamical equations of $\theta$ and $r$ are obtained. Effect of
angular momentum and magnetic field on motion of neutral and charged
particles is observed graphically. It is noticed that a particle
with less angular momentum will approach the black hole more
closely, as compared to the case when angular momentum is large. We
find out the condition on energy of the particle required to escape
or to remain bounded in orbit. Expressions for escape velocity of a
charged particle moving around RN-BH, in the presence of magnetic
field in the vicinity of black hole, are also obtained. Trajectories
of the escaping charged particle are also shown graphically. It is
observed that a slight change in the initial condition of the
colliding particle affects the escaping behavior. Presence of
magnetic field also disturbs the escaping trajectories. Behavior of
effective potentials is studied in details and its effect on the
stability of the orbits is explained graphically. More importantly a
comparison of effective potentials, obtained in the presence and
absence of magnetic field, is established. It is seen that presence
of magnetic field increase the stability of the orbits of the moving
particles, in fact two stable regions (local minima) are obtained in
contrast to the only stable region obtained in the case when
magnetic field is absent. Such analysis helps us to understand the
effect of black hole on its surrounding matter. We intend to  extend
the similar analysis for RN-de-Sitter black hole.
\\
\textbf{Acknowledgement:} The authors would like to thank the
referees for their useful comments to improve this work.


\end{document}